\documentclass{ws-rv9x6}
\usepackage{ws-rv-van,ifpdf,color}
\makeindex

\newcommand{\Ref}[1]{Ref.~[\refcite{#1}]}
\newcommand{\Refs}[1]{Refs.~[\refcite{#1}]}

\newcommand{\Eq}[1]{Eq.~(\ref{#1})}

\newcommand{\Real}{{\, \rm Re \!}}
\newcommand{\intd}[1]{\!\! {\rm d} #1 \,}
\newcommand{\intdd}[1]{\!\! {\rm d^2} #1 \,}

\newcommand{\xcoord}{{\bf x}}
\newcommand{\xring}{x}
\newcommand{\qcoord}{{\bf q}}
\newcommand{\qring}{q}
\newcommand{\tPh}{\tau_{\varphi}}
\newcommand{\tH}{\tau_{H}}
\newcommand{\tTh}{\tau_{\text{Th}}}
\newcommand{\tDw}{\tau_{\textrm{dw}}}
\newcommand{\tT}{{\tau_T}} 
\newcommand{\ETh}{E_{\text{Th}}}
\newcommand{\LDw}{L_{\textrm{dw}}}
\newcommand{\Efield}{{\bf E}}
\newcommand{\cooperon}{{\cal C}}
\newcommand{\decayfunc}{{\cal F}}
\newcommand{\lagrangian}{{\cal L}}

\newcommand{\crw}{\textrm{crw}}
\newcommand{\envelope}{\textrm{en}}
\newcommand{\Tmin}{T_{\text{dil}}}
\newcommand{\Tmax}{T_{\text{ph}}}
\newcommand{\gcont}{g_{\text{cont}}}
\newcommand{\glead}{g_{\text{lead}}}

\begin{document}

\chapter[Dephasing Time in Disordered Mesoscopic Rings]{
         Dimensional Crossover of the Dephasing Time in Disordered
         Mesoscopic Rings: From Diffusive through Ergodic to $0D$
         Behavior}

\author{M. Treiber, O.M. Yevtushenko, F. Marquardt and J. von Delft}

\address{Arnold Sommerfeld Center and Center for Nano-Science,
         Ludwig Maximilians University, Munich, D-80333, Germany}
\author[M.\ Treiber et al.]{I.V. Lerner}

\address{School of Physics and Astronomy, University of Birmingham,
         Birmingham, B15 2TT, UK}

\begin{abstract}
We analyze dephasing by electron interactions in a small disordered
quasi-one dimensional ($1D$) ring weakly coupled to leads, where
we recently predicted a crossover for the dephasing time $\tPh(T)$
from diffusive or ergodic $1D$ ($\tPh^{-1} \propto T^{2/3}, T^{1}$)
to $0D$ behavior ($\tPh^{-1} \propto T^{2}$) as $T$ drops below the
Thouless energy $\ETh$ \cite{2009_our_prb}.
We provide a detailed derivation of our results, based on an influence
functional for quantum Nyquist noise, and calculate all leading and
subleading terms of the dephasing time in the three regimes.
Explicitly taking into account the Pauli blocking of the Fermi sea in
the metal allows us to describe the $0D$ regime on equal footing as
the others.
The crossover to $0D$, predicted by Sivan, Imry and Aronov for
$3D$ systems \cite{1994_sivan_imry_aronov_quantumdot}, has so far
eluded experimental observation.
We will show that for $T \ll \ETh$, $0D$ dephasing governs
not only the $T$-dependence for
the smooth part of the magnetoconductivity but also for the amplitude
of the Altshuler-Aronov-Spivak oscillations, which result only
from electron paths winding around the ring.
This observation can be exploited to filter
out and eliminate contributions to dephasing from trajectories which do
not wind around the ring, which may tend to mask the $T^{2}$ behavior. 
Thus, the ring geometry holds
promise of finally observing the crossover to $0D$ experimentally.
\end{abstract}

\body

\section{Introduction}
  \label{SectionIntroduction}

Over the last twenty{-}five years many theoretical and experimental
works addressed quantum phenomena in mesoscopic disordered metallic
rings \cite{1997_imry_mesobook}.
{This subject was launched in part by several seminal papers by Joe
Imry and his collaborators {
\cite{1983_buettiker_imry_landauer_josephsonbehavior,
1984_buettiker_imry_azbel_quantumoscillations,
1984_gefen_imry_azbel_quantumoscillations,
1985_buttiker_imry_manychannelrings,
1985_murat_gefen_imry_ensembleaveraging,
1986_imry_shiren_energyaveraging,1986_stone_imry_periodicity,
1990_altshuler_gefen_imry_persistentdifferences}}, and continues to be
of great current interest.}
One intensively{-}studied topic involves
persistent currents, which can flow without dissipation due to quantum
interference in rings prepared from normal metals 
\cite{1983_buettiker_imry_landauer_josephsonbehavior,
1990_levy_persistentcurrents,2009_bluhm_persistentcurrents,
2009_bleszynski-jayich_persistentcurrents,2009_imry_tirelesselectrons}.
{Attention was also paid to Aharonov-Bohm oscillations in the
conductance through a mesoscopic ring attached to two leads \cite{1984_buettiker_imry_azbel_quantumoscillations,
1984_gefen_imry_azbel_quantumoscillations,
1985_buttiker_imry_manychannelrings,1985_webb_washburn_aharonovbohm,
1986_stone_imry_periodicity},and the closely related oscillations of the negativeweak localization (WL) correction to the magnetoconductivity \cite{1981_aas_oscillations,1981_sharvin_sharvin}.}These oscillations result from the interference of closed trajectories
which have a non-zero winding number acquiring the Aharonov-Bohm phase.
Both persistent currents and magnetooscillations require the ring to be
phase coherent, since any uncertainty of the quantum phase due to the
environment or interactions immediately suppresses all interference
phenomena \cite{1990_stern_aharonov_imry_phaseuncertainty}.

The mechanism of dephasing in electronic transport and its dependence
on temperature $T$ in disordered conductors was studied in numerous
theoretical \cite{1990_stern_aharonov_imry_phaseuncertainty,
1982_aak_electroncollisions,1983_fukuyama_abrahams_inelastic,
1985_aa_review,1994_sivan_imry_aronov_quantumdot,1999_aag_review,
2004_ludwig_mirlin_ring,2005_texier_montambaux_ring,
2007_marquardt_vondelft_decoherence,2007_vondelft_marquardt_decoherence}
and experimental \cite{1993_marcus_ballisticdot,
1994_yacoby_ballisticdephasing,1995_reulet_bouchiat_mailly_ballistic,
1998_huibers_opendot1,1998_huibers_opendot2,1999_huibers_opendot3,
2000_gougam_relaxation,2008_ferrier_grid} works.
The characteristic time scale of dephasing is called the dephasing time
$\tPh$. At low temperatures phonons are frozen out and dephasing is
mainly due to electron interactions, with the dephasing time $\tPh(T)$
increasing as $T^{-a}$ when $T \to 0 \, , \ a > 0$. 

The scaling of the dephasing time with temperature depends on the
dimensionality of the sample \cite{1982_aak_electroncollisions}. It was
predicted in a pioneering paper by Sivan, Imry and Aronov
\cite{1994_sivan_imry_aronov_quantumdot} that the dephasing time in a
disordered quantum dot shows a dimensional crossover from
$\tPh \propto T^{-1}$, typical for a $2D$ electron gas
\cite{1982_aak_electroncollisions}, to $\tPh \propto T^{-2}$ when the
temperature is lowered into the $0D$ regime:
\begin{equation}\label{0DRegime}
  \hbar/\tPh  \ll  T \ll \hbar/\tTh \, ,
\end{equation}
where $\tTh = {\hbar} / \ETh$ is the Thouless time, i.e. the time
required for an electron to cross (diffusively or ballistically) the
mesoscopic sample; $\ETh$ is the Thouless energy.
{In this low-$T$, 0D regime,} the coherence length and the thermal
length are both larger than the system size, independent of geometry
and real dimensionality of the sample. In this {regime} WL is
practically the only tool to measure the $T$-dependence of dephasing in
mesoscopic wires or quantum dots (the mesoscopic conductance
fluctuations go over to a universal value of order $e^2/h$ for
$T \ll \ETh$ \cite{1997_imry_mesobook}).

Although the $\tPh \propto T^{-2}$ behavior is quite generic, arising
from the fermionic statistics of conduction electrons, experimental
efforts \cite{1998_huibers_opendot1,1998_huibers_opendot2,
1999_huibers_opendot3} to observe it have so far been unsuccessful.
The reasons for this are today still unclear. Conceivably dephasing
mechanisms other than electron interactions were dominant, or the regime
of validity of the $0D$ description had not been reached. In any case,
other ways of testing the dimensional crossover for $\tPh$ are
desirable.

\begin{figure}[t]
\centering
\ifpdf
  \includegraphics[width=0.6\columnwidth]{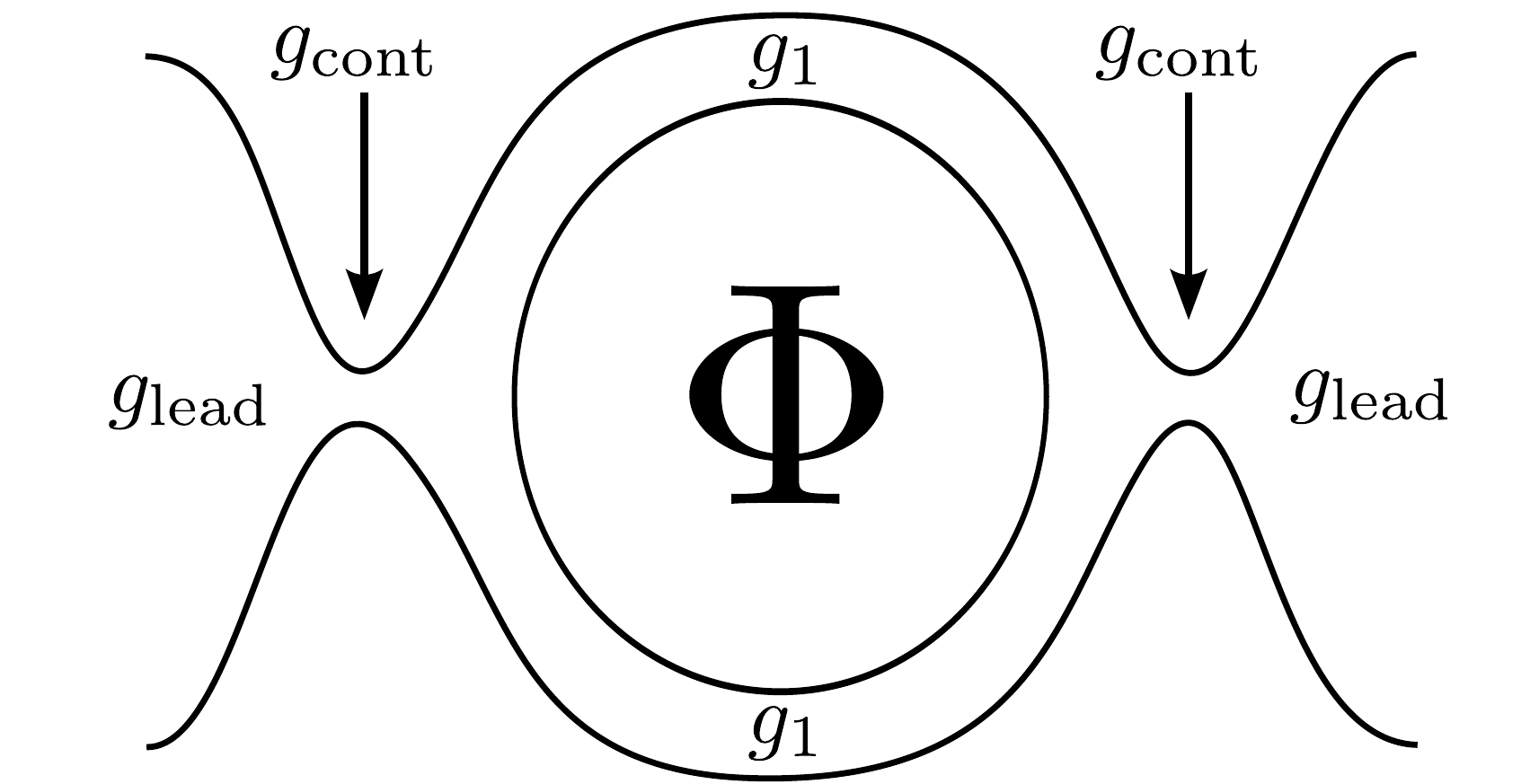}
\else
  \epsfig{file=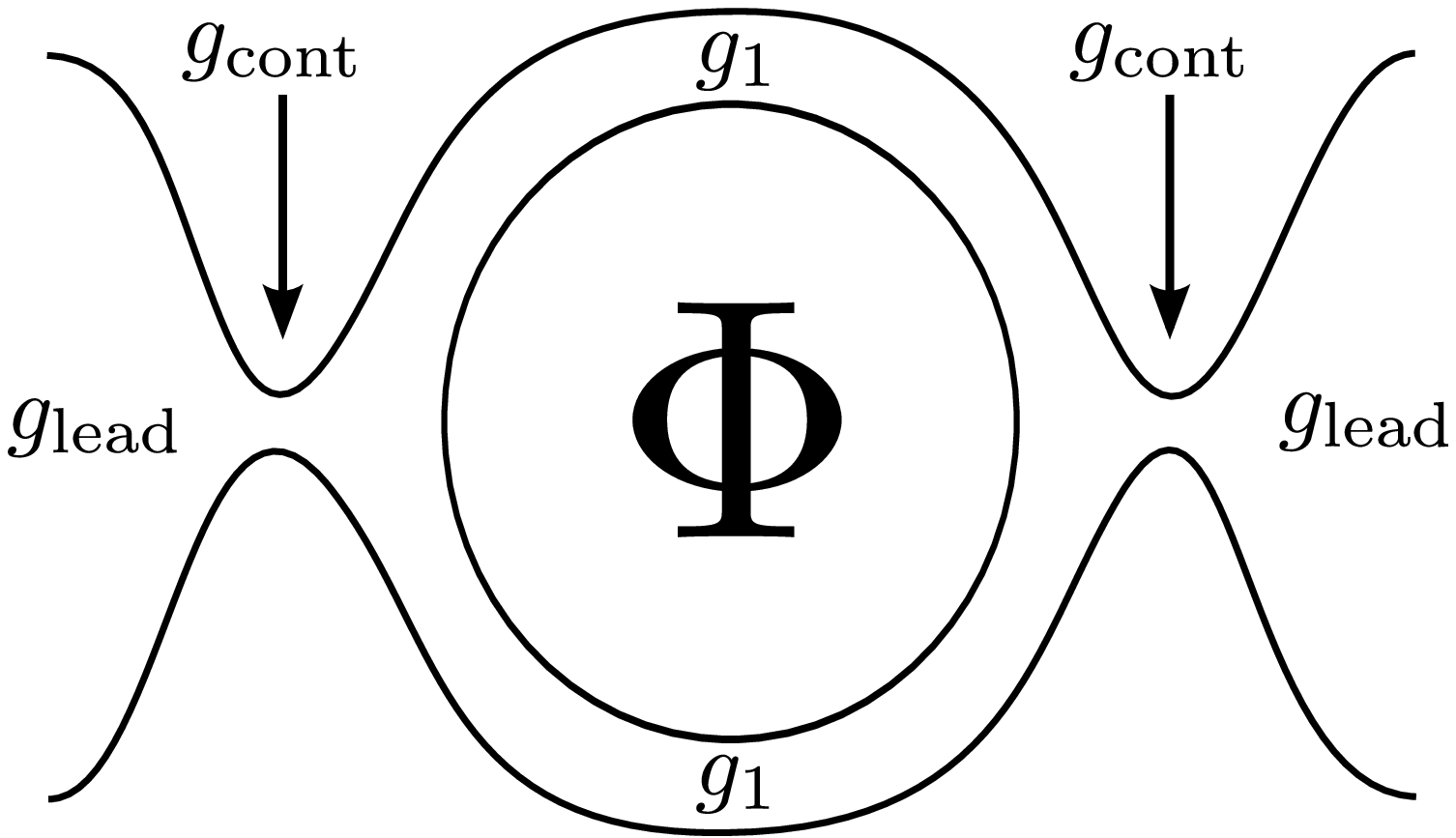,width=0.6\columnwidth}
\fi
\caption{A ring weakly coupled to leads: We assume a metallic system,
         where the conductance at the contacts $\gcont$ is much smaller
         than the conductance of the ring $g_1$ and of the lead $\glead$,
         i.e. $(\glead,g_1) \gg \gcont \gg 1$.
         This assures 
         {(a)} that the average time electrons spend in
         the ring ($\tDw$) is much larger than the average time they
         need to explore the whole {ring} ($\tTh$) and
         {(b)} that the probability for electrons which escaped from the
         ring to return back {to it} is small.}
\label{FigWeakRing}
\end{figure}

In a recent paper \cite{2009_our_prb}, we described the crossover of the
dephasing time to the $0D$ regime in a mesoscopic ring weakly coupled
to leads.
We {considered} a ring of the type shown in Fig.~\ref{FigWeakRing}
with dimensionless $1D$ conductance
\begin{equation}\label{DefConductance}
  g_1 = \frac{h \sigma_0}{e^2} \frac{A}{L} \, ,
\end{equation}
where $A$ and $L$ are the ring's cross section and circumference and
$\sigma_0$ is its classical Drude conductivity.
{In the present paper} we give a detailed derivation of our results
based on an influence functional approach for \emph{quantum noise}. 
This approach explicitly takes into account the Pauli blocking of the
electrons in the metal, which will allow us to describe
quantitatively all regimes of the dephasing time in {a} quasi-$1D$
ring on an equal footing and to calculate first order correction terms
to the dephasing time. In particular, we will see that Pauli blocking
dominates the regime of $0D$ dephasing.
We find that in the $0D$ regime, $T^{-2}$ behavior also emerges for the
amplitude of the Altshuler-Aronov-Spivak (AAS) oscillations of the
conductivity \cite{1981_aas_oscillations} in a magnetic field,
which arise from pairs of time-reversed paths encircling the ring at
least once.
A necessary requirement to reach this regime is that
electron trajectories are effectively confined in the system.
Thus the conductance through the contact, $\gcont$,
is assumed to be much smaller than $g_1$, such that the time an electron
spends inside the ring, the dwelling time $\tDw$, is much larger than
the time an electron needs to {explore} the whole {ring}, i.e.
the Thouless time $\tTh$.

We will show below that after subtracting from the amplitude of the
AAS-oscillations the non-oscillating background, only contributions
to dephasing from paths encircling the ring will contribute.
However, some of these paths may involve loops which not only encircle
the ring, but along the way also enter the lead and reenter the ring
(see Fig.~\ref{FigureRingContributions}(b) below). Such lead-ring
cross-contributions to dephasing will contribute a non-$0D$
$T$-dependence to the conductance and hence tend to mask the $0D$
behavior. We shall argue that by additionally choosing the conductance
of the connected leads, $\glead$, to be larger than $\gcont$, dephasing
due to lead-ring cross-contributions, can be neglected, and the
remaining contributions will be characterized by $0D$ dephasing.

\section{Dephasing and weak localization}
  \label{SectionDephasingWL}

In a disordered metal, the conductivity is reduced by coherent
backscattering of the electrons from impurities, an effect known as
weak localization (WL).
In a semi-classical picture it can be understood as the constructive
interference of closed, time-reversed random-walks through the metal's
impurity landscape{.
It} is most pronounced in systems of low dimensionality $d$ where
the integrated return probability becomes large for long times.
For an infinite system characterized by the diffusion constant
$D=v_F l/d$ ($v_F$ is the Fermi velocity and $l$ is the mean free path),
the probability of a random walk of duration $t$ to return back to its
origin is given by
\begin{equation}\label{Cooperon0}
  \cooperon_0(t) = (4 \pi D t)^{-d/2} \, .
\end{equation}
{To} leading order in $1/g_1$, the relative correction to the
conductance \eqref{DefConductance} can be written {as}
\begin{equation}\label{WLSigma}
  \Delta g = \frac{\Delta \sigma}{\sigma_0}
    = -\frac{1}{\pi \nu} \int_0^{\infty} \intd{t} \cooperon(t)
  \, ,
\end{equation}
where $\nu$ is the density of states per volume in the ring and we
have set $\hbar=1$ henceforth.
The function $\cooperon(t)$ is the so called Cooperon propagator
corresponding to the interference amplitude of the time-reversed random
walks. $\cooperon(t)$ reduces to \Eq{Cooperon0} if time-reversal
symmetry is fully preserved. Processes which destroy this symmetry lead
to a suppression of this contribution at long times, since the random
walks and their time-reversed counterparts acquire a different phase.
The model we are considering assumes a suppression of the Cooperon
of the following form
\begin{equation}\label{CooperonDecay}
  \cooperon(t) \equiv \cooperon_0(t) 
                      \exp\left[ -t/\tH -t/\tDw  - \decayfunc(t) \right]
                      \, .
\end{equation}
In \Eq{CooperonDecay}, we consider dephasing due to the effect of an
external magnetic field leading to the cutoff $\tH \sim 1/H$ of the
integral in \Eq{WLSigma} \cite{1981_altshuler_aronov_magnetic}.
Furthermore, our model of an almost isolated ring assumes an average
dwelling time, $\tDw$, of the electrons in the ring
\cite{1998_mccann_lerner_ucf}.

Our primary interest is the effect of electron interactions, which we
describe in terms of the \emph{Cooperon decay function} $\decayfunc(t)$,
which grows with time and may be used to define a dephasing time via
\begin{equation}
  \decayfunc(\tPh) = 1 \, .
\end{equation}
Dephasing due to electron interactions can be understood roughly as
follows: At finite temperatures the interactions lead to thermal
fluctuations (noise) of the electron's potential energy $V(\xcoord, t)$.
Then the closed paths contributing to WL and their time-reversed
counterparts effectively ``see'' a different local potential, leading to
a phase difference.
This is most clearly seen in a path integral representation of the
Cooperon in a time-dependent potential
\cite{1982_aak_electroncollisions}, which is given by
\begin{equation}
  \cooperon(t) \propto \int_{\xcoord(0)=\xcoord_0}^{\xcoord(t)=\xcoord_0}
                 \mathcal{D}\xcoord \
                 e^{i \varphi{(t)}}
                 e^{- \int_0^t \intd{t_1} \lagrangian(t_1)}
                 \, .
\end{equation}
Here the Lagrangian $\lagrangian(t_1) = \dot{\xcoord}^2(t_1)/4D$
describes diffusive propagation, and $\varphi{(t)}$ is a phase
corresponding to the time-reversed structure of the Cooperon:
\begin{equation}\label{CooperonPhase}
  \varphi{(t)} = \int_0^t \intd{t_1} \left[
              V(\xcoord(t_1), t_1) - V(\xcoord(t_1), t - t_1)
            \right] \, .
\end{equation}
Assuming that the noise induced by electron interactions is Gaussian,
the decay function $\decayfunc(t)$ in \Eq{CooperonDecay}
can be estimated from
$\decayfunc(t) = \frac{1}{2} \langle \overline{\varphi^2} \rangle_{\textrm{crw}}$,
where $\overline{\cdots}$ denotes averaging over realizations of the
noise and $\langle \dots \rangle_{\crw}$ over closed random walks
of duration $t$ from $\xcoord_0$ back to $\xcoord_0$.
$\decayfunc(t)$ is then given in terms of a difference of the noise
correlation functions, taken at reversed instances of time:
\begin{equation}\label{DecayFunc}
  \decayfunc(t) = \int_{0}^{t} \intdd{t_{1,2}}
                  \Bigl\langle
                       \overline{V V} (\xcoord_{12}, t_{12})
                     - \overline{V V} (\xcoord_{12}, \bar t_{12})
                  \Bigr\rangle_{\crw} \, .
\end{equation}
Here $t_{12} = t_1 - t_2$ and $\bar t_{12} = t_1 + t_2 - t${, while}
$\xcoord_{12} = \xcoord(t_1) - \xcoord(t_2)$ is the distance of two
points of the closed random walk taken at times $t_1$ and $t_2$.
{For an infinite wire and the case of classical Nyquist noise
{(defined in \Eq{VVclassical} below)},
\Eq{DecayFunc} has been shown \cite{2007_marquardt_vondelft_decoherence,
2007_vondelft_marquardt_decoherence} to give results practically
equivalent to the exact results obtained in
\Ref{1982_aak_electroncollisions}.}

\section{Thermal noise due to electron interactions}
  \label{SectionThermalNoise}

Electron interactions in the metal lead to thermal fluctuations of the
electric field $\Efield$, {producing} so-called Nyquist noise. In
the high temperature limit, it can be obtained from the classical
Fluctuation-Dissipation Theorem leading to a field-field
correlation function in $3$D of the form
\begin{equation}\label{EECorrelatorClassic}
  \overline{\Efield \Efield} (\qcoord, \omega)
    \xrightarrow{{|\omega|} \ll T} \frac{2 T}{\sigma_0} \, .
\end{equation}
Note that the fluctuations of the fields do not depend on $\qcoord$ or
$\omega$, i.e. they correspond to white noise in space and time.  To
describe dephasing in a quasi-$1D$ wire, we need the correlation
function of the corresponding scalar potentials $V$ in a quasi-$1D$
wire.  {Since $E = \frac{1}{e} \nabla V$, the noise correlator that
  corresponds to the classical limit \eqref{EECorrelatorClassic} has
  the form\begin{equation}\label{VVclassical} \overline{V V}_{\rm
      class} (\qcoord, \omega) = \frac{2 T e^2}{\sigma_0}
    \frac{1}{\qcoord^2} \, .\end{equation}This so-called classical
  Nyquist noise is frequency independent, i.e.  corresponds to ``white
  noise''.}  {For present purposes, however, we need its
  generalization to the case of quantum noise, valid for arbitrary
  ratios of ${|\omega|}/T$.  In particular, $\overline{VV}$ is
  expected to become frequency-dependent: it should go to zero for
  ${|\omega|} \gg T$, since the Pauli principle prevents scattering
  processes into final states occupied by other electrons in the Fermi
  sea \cite{2002_imry_decoherenceatzero}. A careful analysis of
  quantum noise has been given recently} in
\Ref{2007_marquardt_vondelft_decoherence} and
\Ref{2007_vondelft_marquardt_decoherence}. The authors derived an
effective correlation function for the quantum noise potentials {that
  properly accounts for the Pauli principle. It is} given by
\begin{equation}\label{Vcorrelator0}
  \overline{V V} (\qcoord, \omega)
    =  {\rm Im}{\cal L}^R(\qcoord,\omega) \,
      \frac{\omega/2T}{\sinh\left(\omega/2T\right)^2}
\end{equation}
with
\begin{equation}\label{RetardedInteraction}
  {\cal L}^R(\qcoord,\omega)
  = - \frac{D\qcoord^2 - i\omega}
           {2 \nu D \qcoord^2 + (D\qcoord^2 - i\omega)/V(\qcoord)} \, ;
\end{equation}
$V(q)$ is the Fourier{-}transformed bare Coulomb potential (not
renormalized due to diffusion) in the given effective dimensionality.

If the momentum and energy transfer which dominates dephasing is small
then the second term of the denominator of \Eq{RetardedInteraction} can
be neglected so that \Eq{RetardedInteraction} reduces to
\begin{equation}\label{ImLSimple}
  {\rm Im}{\cal L}^R(\qcoord,\omega)
    \simeq \frac{\omega}{2 \nu D \qcoord^2} \, .
\end{equation}
This simplification holds true, in particular, in the high temperature
(diffusive) regime where $\omega \ll T$
\cite{1982_aak_electroncollisions}. We will {argue below (see Eqs.
\eqref{UnitaryLimitDenominator} to \eqref{CoulombPot})} that the same
simplification can be used in the low temperature regime where
${|\omega|} \sim T \lesssim \ETh $ \cite{2009_our_prb}.

Inserting \Eq{ImLSimple} in \Eq{Vcorrelator0} with
$\sigma_0 = 2 e^2 \nu D / A$, where $A$ is the cross-section
perpendicular to the current direction, we obtain
\begin{equation}\label{VCorrelatorFourier}
  \overline{V V} (\qcoord, \omega) = \frac{2 e^2 \, T}{\sigma_0 A}
                               \, \frac{1}{\qcoord^2}
                               \, \left(\frac{\omega / 2T}
                                             {\sinh(\omega / 2T)}
                                  \right)^2 \, .
\end{equation}

In the time and space domain, this correlator factorizes into a product
of time- and space-dependent parts:
\begin{equation}\label{Vcorrelator}
  \overline{V V} (\xcoord, t) = \frac{2 e^2 \, T}{\sigma_0 A}
                               \, Q(\xcoord) \, \delta_T(t) 
  \, ,
\end{equation}
where $\delta_T(t)$ is a broadened delta function of width $1/T$ and
height $T$: 
\begin{equation}\label{DeltaT}
  \delta_T(t) =  \pi T w(\pi \, T \, t) \, , 
  \qquad
         w(y) = \frac{y \coth(y) - 1}{\sinh^2(y)} \, .
\end{equation}
{The fact that the noise correlator \eqref{Vcorrelator} is
proportional to a \emph{broadened} peak {$\delta_T(t)$} is a direct
consequence of the effects of Pauli blocking. Previous approaches often
used a sharp Dirac-delta peak instead. In the frequency domain this
corresponds to white noise and leads to \eqref{VVclassical}, instead of
our frequency-dependent form \eqref{Vcorrelator0}.} Such a
``classical'' treatment {reproduces correct} results for the
dephasing time when processes with small energy transfers
${|\omega|} \ll T$ dominate. {However, it} has been shown in
\Ref{1994_sivan_imry_aronov_quantumdot} that this is in fact not the
case in the $0D$ limit $T \ll \ETh${, where} the main contribution
to dephasing is due to processes with ${|\omega|} \simeq T${.
Thus, the results become dependent on the form of the cutoff that
eliminates modes with ${|\omega|} > T$ to account for the Pauli
principle. For such purposes, previous treatments typically introduced a
sharp cutoff, $\theta(T - {|\omega|})$, by hand. However, the
precise} form of the cutoff becomes important in an analysis interested
not only in qualitative features, but quantitative details.
{The virtue of (1.11) is that it encodes the cutoff in a
quantitatively reliable fashion. (For example, it was shown
\cite{2007_marquardt_vondelft_decoherence} to reproduce
a result first obtained in \Ref{1999_aag_review}, namely the subleading
term in an expansion of the large-field magnetoconductance (for quasi-1D
wires) in powers of the small parameter $1/\sqrt{T\tau_H}$.)}

The position-dependent part of \Eq{Vcorrelator}, the so-called diffuson
at zero frequency $Q(\xcoord)$, is the time-integrated solution
of the diffusion equation.
In the isolated system, it satisfies
\begin{equation}\label{QDiffEq}
  - \Delta Q(\xcoord) = \delta(\xcoord) \, ,
\end{equation}
with given boundary conditions, which govern the distribution of the
eigenmodes of $Q$. In an isolated system, where a $\qcoord=0$ mode is present,
$Q(\xcoord)$ diverges. However, the decay function is still regular,
since terms in $Q$ which do not depend on $\xcoord$ simply cancel out
in \Eq{DecayFunc} and cannot contribute to dephasing.

To evaluate the decay function \Eq{DecayFunc}, we note that only the
factor $Q(\xcoord)$ in \Eq{Vcorrelator} depends on $\xcoord$, thus,
the average $\langle Q(\xcoord) \rangle_{\crw}$ has to be calculated.
This will be done in the next section for an almost isolated ring.
Then, after a qualitative discussion in \ref{SectionQualitativePicture},
we proceed by evaluating $\decayfunc(t)$ in section
\ref{SectionAnalyticalResults}.

\section{Diffusion in the almost isolated ring}
  \label{SectionDiffusionRing}

\begin{figure}[t]
\centering
\ifpdf
  \includegraphics[width=\columnwidth]{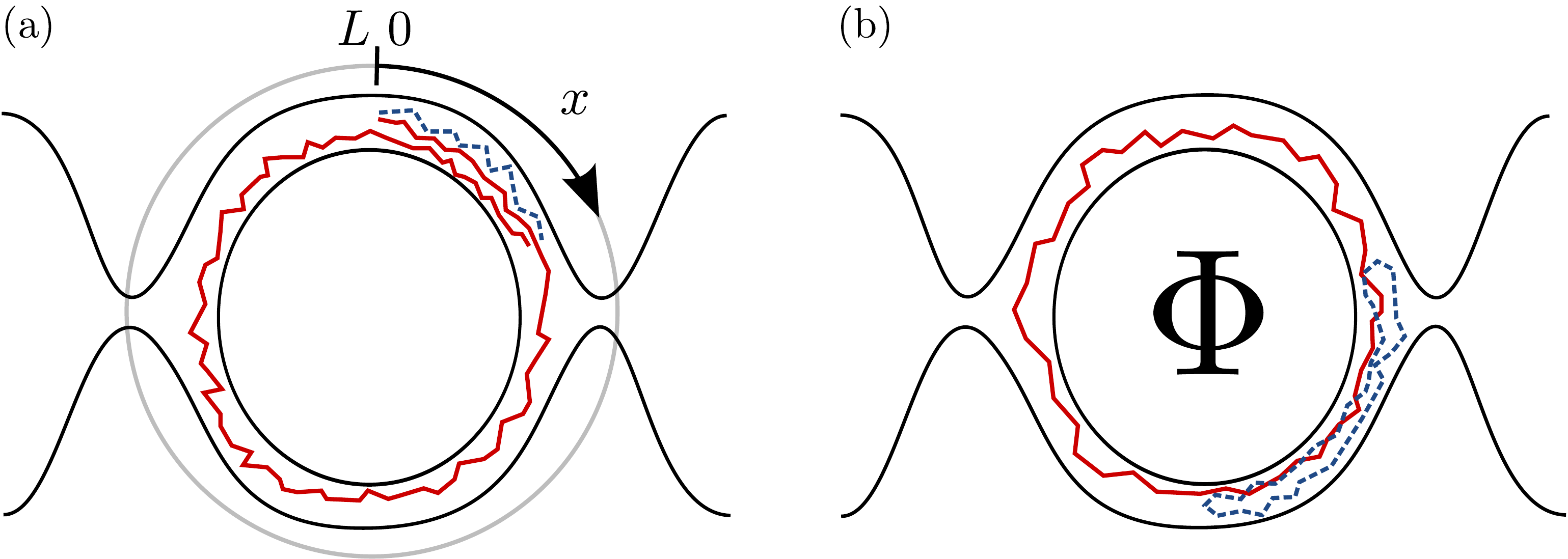}
\else
  \epsfig{file=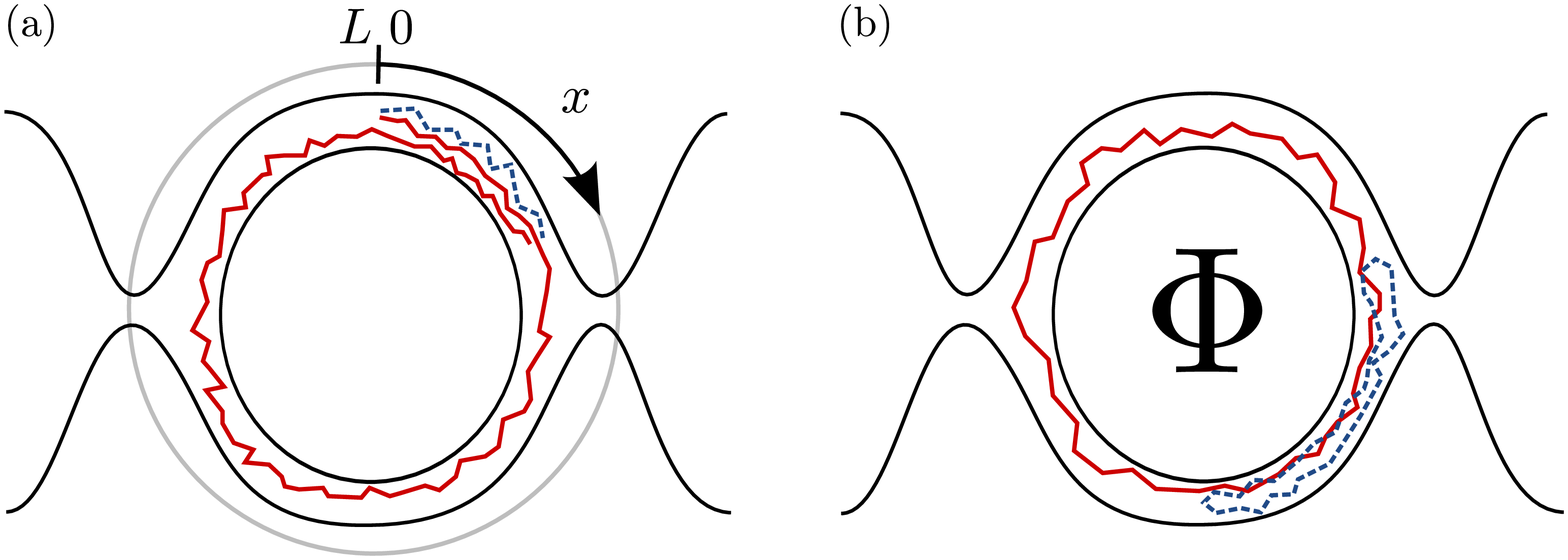,width=\columnwidth}
\fi
\caption{(a) Illustration of our choice of the coordinate system:
         Both paths have the same start and end point
         ($0 \mapsto \xring$), but the dashed path has
         winding number $n=0$ and the solid path $n=1$.
         (b) Two \emph{closed} paths in the ring contributing to the
         Cooperon. The contribution of the solid path
         (with winding number $n=1$) is affected by the flux $\Phi$,
         since the path (and it's time-reversed counterpart) acquire
         an Aharonov-Bohm phase when interfering with itself at their
         origin. This gives rise to the Altshuler-Aharonov-Spivak
         oscillations. The dashed path with $n=0$ is unaffected
         by the flux, since the acquired phase at the origin is zero.}
\label{FigWeakRingAAS}
\end{figure}
\noindent
The probability density of a random walk in a $1D$, infinite, isotropic
medium to travel the distance $\xring$ in time $t$ is given by
\begin{equation}\label{FreeDiffuson}
  P_0(\xring, t) = \frac{1}{\sqrt{4 \pi D t}} \,
                    e^{ -\xring^2/4 D t} \, .
\end{equation}
In an isolated ring, electrons can reach each point without
or after winding around the ring $n$ times, where $n$ is called
\emph{winding number}.
Denoting the probability density for the latter type of path by
$P_n(\xring,t)$, the diffusion probability density can be expanded
in $n$ as
\begin{equation}\label{PropWinding}
  P(\xring, t) = \sum_{n=-\infty}^{+\infty} P_n(x, t) \, ,
  \quad
  P_n(x, t) = \frac{1}{\sqrt{4 \pi D t}} \,
              e^{ -(\xring+nL)^2 / 4 D t} \, ,
\end{equation}
where $L$ is the circumference of the ring and $\xring \in [0,L]$ is
the cyclic coordinate along the ring, see Fig.~\ref{FigWeakRingAAS}(a).
To model the effect of the two contacts of the ring, we assume that
an electron, on average, stays inside the ring only for the duration of
the dwelling time $\tDw$, introduced in \Eq{CooperonDecay}, and then
escapes with a vanishing return probability. This simplified model of
homogeneous dissipation, strictly applicable only in the limit
$\tTh \ll \tDw$ and for a very large lead conductance, captures all the
essential physics of the $0D$ crossover we are interested in.
Our present assumptions lead to the following replacement of the
diffusion probability density:
\begin{equation}\label{PropReplaceDwelling}
  P(\xring, t) \to P(\xring, t) \ e^{-t/\tDw} \, .
\end{equation}

Furthermore, the spatial dependence of the noise correlation function
\eqref{RetardedInteraction} acquires an additional dissipation term in
the denominator. Thus, in contrast to the isolated case, $Q(\xring)$
now satisfies the Laplace transform of the diffusion equation, given by
\begin{equation}
  \left[ \frac{1}{\LDw^2} - \Delta \right] Q(\xring)
    = \delta(\xring) \, ,
\end{equation}
where $\LDw = \sqrt{\tDw D}$.
For a ring with circumference $L$ we obtain
\begin{equation}\label{RingDiffuson}
  Q(\xring) = \frac{\LDw}{2} 
              \frac{\cosh\Bigl([L-2|\xring|]/2 \LDw\Bigr)}
                   {\sinh(L/2 \LDw)} \, .
\end{equation}
We can expand \Eq{RingDiffuson} for the almost isolated ring in powers
of $\tTh/\tDw \ll 1$:
\begin{equation}\label{QAlmostIsolated}
  Q(\xring) \approx C - \frac{|\xring|}{2}\left(1 - \frac{|\xring|}{L} \right)
  + {\cal O}\left( \frac{\tTh}{\tDw} \right) \, ,
\end{equation}
where {the $x$-independent first term, $C = L \tDw / \tTh$,
describes the contribution of the zero mode}. As
expected, see the discussion after \Eq{QDiffEq}, it diverges in the
limit $\tTh/\tDw \to 0$.

Having described the diffuson in our model of the almost isolated ring,
we can proceed by calculating the closed random walk average ($\crw$) of
\Eq{QAlmostIsolated}. We will see {below that} we need to consider
the random walk average with respect to closed paths with a specific
winding number $n$ only. For an isolated ring, using \Eq{PropWinding},
it can be written as
\begin{equation}\label{RwAverage}
  \langle Q \rangle_{\crw}(t_{12}, n)
    = \int_0^L \intdd{\xring_{1,2}} Q(\xring_{12})
                               P_{\crw}(\xring_{12}, t_{12}, n) \, {,}
\end{equation}
with
\begin{equation}\label{DiffusionRingAveraging}
  P_{\crw}(\xring_{12}, t_{12}, n)
    = \!\!\!\! \sum_{i+j+k=n} \!\!\!\!
      \frac{P_i(\xring_{01},t_1) P_j(\xring_{12},t_{21}) P_k(\xring_{20},t-t_2)}
           {P_n(0,t)} \, ,
\end{equation}
where we used the notation
$\xring_{\alpha \beta} = \xring_{\alpha} - \xring_{\beta}$ and
$t_{\alpha \beta} = t_{\alpha} - t_{\beta}$.
Obviously, the replacement \eqref{PropReplaceDwelling} does not affect
this averaging procedure, so that it remains valid in our model of
homogeneous dissipation.
Note that the expression \eqref{DiffusionRingAveraging} depends in fact
only on $x_{12}$ and not on $x_0$, as can be verified by integrating
both sides of the equation over $x_0$ using the following semi-group
property in the ring:
\begin{equation}
  \int_0^L \intd{\xring_2} P_l(\xring_{12}, t_1)
                            P_m(\xring_{23}, t_2)
    = P_{l+m}(\xring_{13}, t_1 + t_2) \, .
\end{equation}
Doing the average of \Eq{RingDiffuson} according to \Eq{RwAverage}, we
finally obtain
\begin{equation}\label{Qaveraged}
  \langle Q \rangle_{\crw}(t_{12}, n)
    = C - \frac{L}{2} \sum_{k=1}^{\infty}
      \frac{\cos(2 \pi k n u)}{(\pi k)^2}
      e^{- (2 \pi k)^2 \, \ETh \, t_{12} (1-t_{12}/t)} \, .
\end{equation}
It follows that a finite dissipation rate does not affect the decay
function to leading order in $\tTh/\tDw$.

For the Cooperon, an expansion similar to \Eq{PropWinding} can be done.
In addition to that, the dependence of the Cooperon on an external
magnetic field changes due to the ring geometry.
It not only leads to the suppression of the Cooperon at long times, but
also, due to the Aharonov-Bohm effect, to Altshuler-Aronov-Spivak
oscillations \cite{1981_aas_oscillations} of the WL-correction, see
Fig.~\ref{FigWeakRingAAS}(b). Combining these remarks with
\Eq{CooperonDecay} and inserting \Eq{Cooperon0} with $d=1$, we write
the Cooperon in our model as
\begin{equation}\label{CooperonSum}
  \cooperon(t) = \sum_{n=-\infty}^{+\infty}
                 \frac{e^{-(nL)^2/4 D t}}{\sqrt{4 \pi D t}}
                 e^{ -t/\tH -t/\tDw -\decayfunc_n(t)}
                 e^{i n \theta } \, ,
\end{equation}
where $\theta = 4 \pi \phi / \phi_0$
and $\phi = \pi (L/2\pi)^2 H$ is the flux through the ring
($\phi_0 = 2 \pi c/e$ is the flux quantum).
Note that the decay function $\decayfunc$ is now a function of $n$:
Since we used an expansion in winding numbers $n$, we should consider
the phase \eqref{CooperonPhase} acquired by paths with the winding
number $n$ only. Thus, the $\crw$-average in \Eq{DecayFunc} has to
be performed with respect to paths with given winding number $n$ only,
as anticipated in \Eq{RwAverage}.

\section{Qualitative picture from the perturbative expansion of the Cooperon}
  \label{SectionQualitativePicture}

\begin{figure}[t]
\centering
\ifpdf
  \includegraphics[width=0.5\columnwidth]{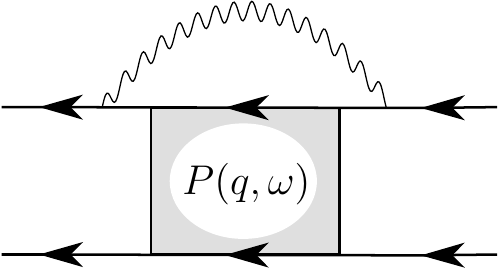}
\else
  \epsfig{file=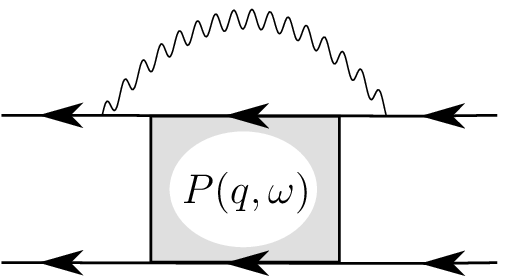,width=0.5\columnwidth}
\fi
\caption{Typical diagram from the expansion of the Cooperon self energy.
         The shaded area denotes impurity lines, described by 
         the diffusion propagator \Eq{DiffusonFourier}. The upper solid
         lines correspond to a retarded electron Green's function and
         the lower to an advanced (or vice versa). Wiggly lines denote
         electron interactions, described by \Eq{Vcorrelator}.}
\label{FigCoopSelfDiag}
\end{figure}

In our previous paper \cite{2009_our_prb} we showed how all the regimes
of the dephasing time in an isolated ring can be understood
qualitatively from the influence functional picture. In particular,
we demonstrated how $0D$ dephasing emerges from the assumption of
{a noise field that is effectively "frozen" on the time scale
$\tau_{\rm Th}$ (since ${|\omega|} \simeq T \ll \ETh$), }
leading to a drastically reduced dephasing rate.
The qualitative behavior of $\tPh$ also follows from the standard
perturbative expression for the Cooperon self-energy.
Such self-energy diagrams are of the type shown in
Fig.~\ref{FigCoopSelfDiag} and were first evaluated in
\Ref{1983_fukuyama_abrahams_inelastic}. This diagram and its
complex conjugate give contributions to the dephasing time of the form
\begin{equation}\label{TPhiPertTheoryNoCutoff}
  \frac{1}{\tPh} \propto \int \intd{\omega}
                         \int \intd{\qring}
                         \overline{V V}(\qring, \omega)
                         \Real\left[P(\qring, \omega)\right]
  \, ,
\end{equation}
where the diffuson $P(\qring, \omega)$ is given by the Fourier transform
of \Eq{FreeDiffuson}:
\begin{equation}\label{DiffusonFourier}
  P(\qring, \omega) = \frac{1}{D\qring^2 - i\omega} \, .
\end{equation}
We {have} already {mentioned} that large energy transfers are
suppressed according to \Eq{VCorrelatorFourier} leading to an upper
cutoff at $T$ of the frequency {integration.
Furthermore}, it was shown in \Refs{2007_marquardt_vondelft_decoherence,
2007_vondelft_marquardt_decoherence} that vertex contributions to
these self-energy diagrams cure the infrared divergences in the frequency
integration{,} leading to a cutoff at $1/\tPh$. Such fluctuations are
simply too slow to influence the relevant paths.
Note that in contrast to the perturbative treatment presented in this
section, the path integral calculation leading to the expression
\Eq{DecayFunc} for the decay {function is} free of these IR divergences.
In fact, it was shown that the first term of \Eq{DecayFunc}
corresponds, when compared to a diagrammatic evaluation of the Cooperon
self-energy, to the so-called self-energy contributions
(shown in Fig.~\ref{FigCoopSelfDiag}), while the second term corresponds
to the so-called vertex contributions.  

In the ring geometry, the diffuson has quantized momenta and the $q=0$
mode can not contribute. For a qualitative discussion we may take this
into account by inserting a lower cutoff $1/L$ of the momentum
integration.

Taking into account the above remarks, we can estimate the dephasing
time as
\begin{equation}\label{TPhiPertTheory}
  \frac{1}{\tPh} \propto \frac{T}{g_1 L}
                         \int_{1/\tPh}^{T} \intd{\omega}
                         \int_{1/L}^{\infty} \intd{\qring}
                         \frac{D}{(D\qring^2)^2 + \omega^2}
  \, .
\end{equation}
\Eq{TPhiPertTheory} illustrates succinctly that the modes
dominating dephasing lie near the infrared cutoff
($\omega \simeq \tPh^{-1}$ or $\ETh$)
for the diffusive or ergodic regimes, but near the ultraviolet cutoff
$\omega \simeq T$ for the $0D$ regime,
which is  why, in the latter, the broadening of $\delta_T (t)$ becomes
important.
Performing the integrals in Eq. (1.28) and solving for $\tPh$
selfconsistently, we find three regimes:
\begin{enumerate}
  \item The \emph{diffusive regime}{,} for $ \tT \ll \tPh \ll \tTh ${,} with
        \begin{equation}
          \tPh \propto (g_1 / \sqrt{\ETh}T)^{2/3} \, ;
        \end{equation}
  \item the \emph{ergodic regime}{,} for $ \tT \ll \tTh \ll \tPh${,} with
        \begin{equation}
          \tPh \propto g_1 / T \, ;
        \end{equation}
  \item and the \emph{$0D$ regime}{,} reached at $\tTh \ll \tT \ll \tPh${,} with
        \begin{equation}\label{Perturb0DRegime}
          \tPh \propto g_1 \ETh/T^2 \, . 
        \end{equation}
\end{enumerate}
Here, $\tT = \sqrt{D/T}$ is the thermal time.
{Expressing \eqref{Perturb0DRegime} in terms of the level spacing
$\delta = \ETh / g_1$ we find $\tPh \delta \propto \ETh^2/T^2$. This
ratio is $\gg 1$ in the 0D regime, implying that dephasing is so weak
that the dephasing rate $1/\tPh$ is smaller than the level spacing.}

\section{Results for the Cooperon decay function}
  \label{SectionAnalyticalResults}

\begin{figure}[t]
  \ifpdf
    \includegraphics[width=\columnwidth]{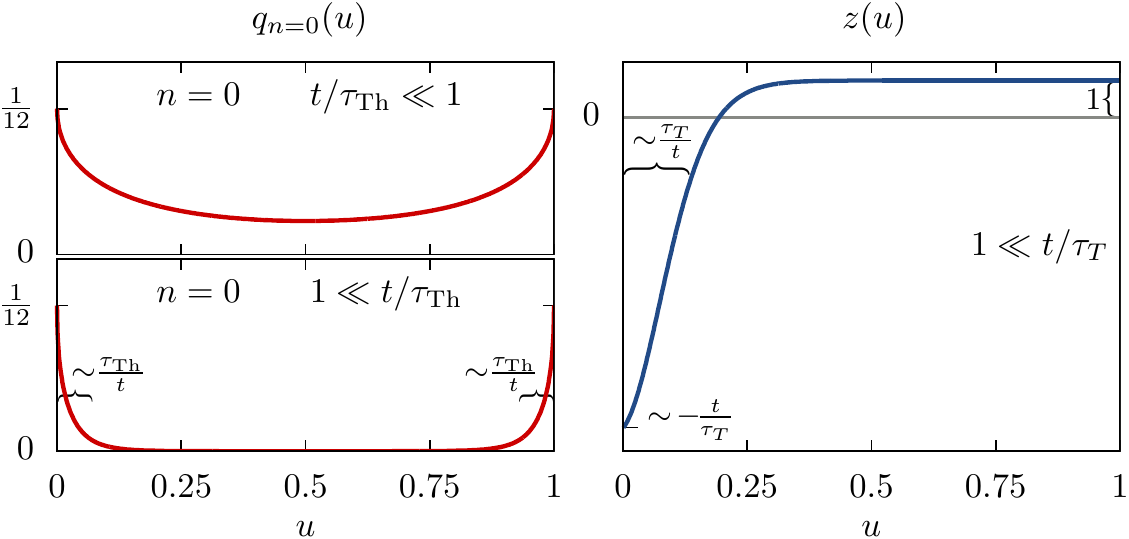}
  \else
    \epsfig{file=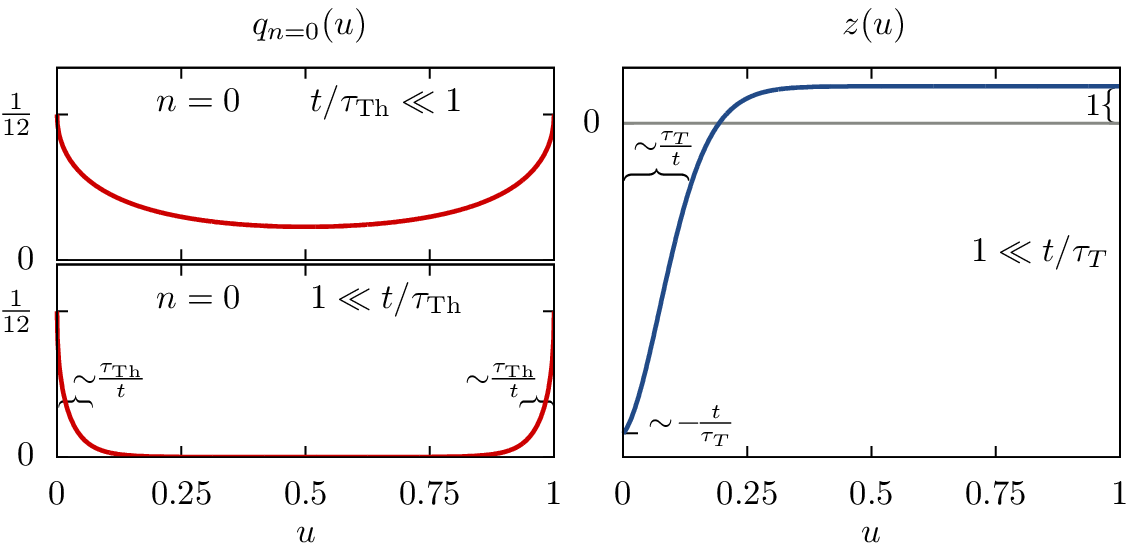,width=\columnwidth}
  \fi
  \caption{Functions $q_{n=0}(u)$ (left panel) defined in \Eq{Dimlessq}
                      and $z(u)$ (right panel) defined in \Eq{Dimlessz}.}
  \label{FigZq}
\end{figure}
For a systematic analysis of the Cooperon decay function, we rewrite
\Eq{DecayFunc} in terms of an integral over the dimensionless variable
$u = t_{12}/t$:
\begin{equation}\label{DecFuncFinal}
  {\decayfunc}_n(t) = \frac{4 \pi T t}{g_1}
                      \int_{0}^{1} \intd{u}
                      z(u) \, q_n(u) \, ,
\end{equation}
where the kernel
\begin{equation}\label{Dimlessz}
  z(u) = - 2 \pi T t \, (1 - u) \, w(\pi T t u)
         + \int_{-\pi T t u}^{\pi T t u} \intd{v} w(v)
\end{equation}
depends on the parameter $tT = t/\tT$, and the dimensionless
$\crw$-averaged diffuson
\begin{equation}\label{Dimlessq}
  q_n(u) = \frac{\langle Q \rangle_{\crw}(u t) - C}{L}
       = -\frac{1}{2} \sum_{k=1}^{\infty}
         \frac{\cos(2 \pi k n u)}{(\pi k)^2}
         e^{- (2 \pi k)^2 \,  (t/\tTh) \, u(1-u)} \, ,
\end{equation}
depends on $t/\tTh$, see \Eq{Qaveraged}.
Note that we can add or subtract an arbitrary number from $q_n(u)$
without changing the result, since constant terms in $q_n(u)$,
describing the zero mode, do not contribute to dephasing,
because of the following property of $z(u)$:
\begin{equation}\label{IntegralzC}
  \int_0^1 \intd{u} z(u) = 0 \, .
\end{equation}
Both functions, \Eq{Dimlessz} and \Eq{Dimlessq}, are illustrated in
Fig~\ref{FigZq} in all relevant limiting cases.
Note that in the regime of WL we always have $\tT \ll t$.
In the opposite regime the interaction correction to the conductivity
(Altshuler-Aronov correction) originating from the Friedel
oscillations dominate \cite{1980_aal_interactioneffects}, which we do
not consider here.

We proceed with an asymptotic evaluation of \Eq{DecFuncFinal}.
For large $t/\tTh$, $q_n(u)$ is non-zero ($\simeq \frac{1}{12}$)
only in the intervals $0<u<\tTh/t$ and $1-\tTh/t<u<1$, see
Fig.~\ref{FigZq}. For small $t/\tTh$ and $n=0$ we can use the expansion
\begin{equation}\label{Expansionq}
  q_{n=0}(u) \approx \frac{1}{12} - \frac{1}{\sqrt{\pi}}
                              \sqrt{\frac{t}{\tTh} (u(1-u))}
                            + \frac{t}{\tTh} (u(1-u)) \, .
\end{equation}
For larger $n$ the exponential function in \Eq{Dimlessq} can be expanded
since the sum converges at $k \simeq 1$.

For $\tT \ll t$, $z(u)$ is large ($\sim -t/\tT$) in the interval
$0<u<\tT/t$ and $z(u) \approx 1$ otherwise. Thus, it will be
convenient to decompose $z(u) = \overline{z} + \tilde{z}(u)$ into a
constant part $\overline{z} = +1$ and a peaked part
$\tilde{z}(u) = z(u)-1$.
For contributions of the peaked type one observes that
\begin{equation}\label{Integralz}
  \int_0^1 \intd{u} \tilde{z}(u) \, u^s = \begin{cases}
    -1, \qquad
    & s=0 \, ; \\
    -\sqrt{\frac{\tT}{t}} \frac{\sqrt{2 \pi}}{4}
      |\zeta\left( \frac{1}{2} \right) \! |, \qquad
    & s=1/2 \, ; \\
    -\frac{\tT}{t}, \qquad
    & s = 1 \, .
  \end{cases}
\end{equation}
We identify the following 3 regimes:

\paragraph{Diffusive regime $\tT \ll t \ll \tTh$ and $n=0$:}
Here we can use the expansion \Eq{Expansionq}.
The constant term does not contribute, due to \Eq{IntegralzC}.
The main contribution to the integral comes from values of $u$ where
$z(u) \approx 1$. Thus, we decompose
$z(u) = \overline{z} + \tilde{z}(u)$ as suggested above.
The leading result and corrections
$\propto \sqrt{t/\tTh}$ due to the second and third term in
\Eq{Expansionq} stem from $\overline{z}$. Corrections
$\propto \sqrt{t/\tT}$ can be calculated with the help of
\Eq{Integralz} with $s=1/2$ from the $\tilde{z}(u)$ part.
In total we obtain for $n=0$:
\begin{equation}
  \decayfunc_{n=0}(t) = \frac{\pi^{3/2} \sqrt{\ETh}}{2 g_1} \ T t^{3/2}
  \left( 1 
    + \frac{2^{3/2} \zeta\!\left(\frac{1}{2}\right)}{\pi} \frac{1}{\sqrt{tT}}
    - \frac{4}{3 \sqrt{\pi}} \sqrt{\frac{t}{\tTh}}
  \right) \, .
\end{equation}

\paragraph{Diffusive regime $\tT \ll t \ll \tTh$ and $|n|>0$:}
For winding numbers larger than zero, we expand the exponential function
in \eqref{Dimlessq}.
In contrast to the case of $n=0$, the leading result comes here from
the peaked part $\tilde{z}(u)$. After expanding the exponential
function and doing the sum over $k$, we can apply \Eq{Integralz} with
$s=0$ and $s=1$ to find the leading result and a correction
{$\sim \tT/t$}. For $\overline{z}$, we observe that the first term
vanishes since the integral is over $n$ full periods of $\cos$.
The second term of the expansion gives a correction $\sim t/\tTh$
and in total for $0 < |n| \ll t/\tT$:
\begin{equation}\label{DetailDiffusiveHighWinding}
  \decayfunc_n(t) = {\frac{\pi}{3 g_1} \ T t}
  \left( 1 
    - \frac{2}{n^2} \frac{t}{\tTh}
    - \frac{6}{\pi n} \frac{\tT}{t}
  \right) \, .
\end{equation}
Note that in the diffusive regime, winding numbers $|n|>0$ only
contribute weakly to the conductivity, see \Eq{CooperonSum}.

\paragraph{Ergodic regime $\tT \ll \tTh \ll t$:}
In this regime, the main contribution to the conductivity will not
depend on $n$, since we may neglect the $\cos$-term of $q_n(u)$ as
long as $|n| \ll t/\tTh$. This restriction on $n$ is justified by
the fact that large values of $|n|$ give contributions smaller by
a factor of $\sim \exp(-n^2 t/\tTh)$, see \Eq{CooperonSum}.

Again, we decompose $z(u) = \overline{z} + \tilde{z}(u)$. For the
$\tilde{z}(u)$ part, we use the expansion of $q_{n=0}(u)$,
\Eq{Expansionq}, where the constant term $1/12$ will yield the main
result. Corrections due to the second term of \Eq{Expansionq} are
$\sim \sqrt{\tT/\tTh}$, because of \Eq{Integralz} with $s=1/2$.
For $\overline{z}$, we do the integral over $u$ directly using
$\int_0^1 \intd{u} \exp(-x u (1-u)) \xrightarrow[]{x \rightarrow \infty} 2/x$.
From this we obtain a correction $\sim \tTh/t$ and in total
\begin{equation}\label{DetailErgodic}
  \decayfunc_n(t) = \frac{\pi}{3 g_1} \ T t
  \left( 1
   - \frac{6}{\sqrt{2 \pi}} \sqrt{\frac{\tT}{\tTh}}
   - \frac{1}{30} \frac{\tTh}{t}
  \right) \, .
\end{equation}
It is not surprising that the case {$|n|>0$} in the diffusive regime
gives{, to leading order,} the same results as all $n$ of the
ergodic regime {(compare \Eq{DetailErgodic} to
\eqref{DetailDiffusiveHighWinding}),} since higher winding numbers are
by definition always ergodic: The electron paths explore the system
completely.

\paragraph{$0D$ regime $\tTh \ll \tT \ll t$:}
In this regime, $q_n(u)$ is more sharply peaked than $z(u)$, since
$\tT/t \gg \tTh/t$. This means that the electron reaches the fully
ergodic limit (where $q(u) = {\rm const}$ and no dephasing can occur)
before the fluctuating potential changes significantly.
Thus, the potential is effectively frozen and only small statistical
deviations from the completely ergodic limit yield a phase difference
between the two time-reversed trajectories.
The width of the peak of $z(u)$ becomes unimportant, instead, we
can expand $z(u)$ around $u=0$ and $u=1$. Furthermore, we can expand
the argument of the exponential function in $q_n(u)$ and then extend
the integral to $+\infty$ and scale $u$ by $k \pi$:
\begin{equation}
  \decayfunc_n(t) = \frac{4 \pi T t}{g_1} \int_0^{\infty} \intd{u}
    \left[ \frac{2 \pi T t}{3} -1 -\frac{4 \pi^3}{15} (tT)^3 u^2 \right]
    \sum_{k=1}^{\infty} \frac{\cos(2nu)}{(k \pi)^3}
    e^{- 4 k \pi \ETh t u}
\end{equation}
(the $-1$ in the integrand stems from the region $u \approx 1$).
Now, assuming $|n| \ll t/\tTh$, the integral over $u$ can be done and
then the sum over $k$ evaluated. The result is
\begin{equation}
  \decayfunc_n(t) = \frac{\pi^2 \tTh}{270 \, g_1} \ T^2 t
  \left( 1
   - \frac{3}{2 \pi} \frac{1}{T t}
   - \frac{\pi^2}{210} \frac{T^2}{\ETh^2}
  \right) \, .
\end{equation}
Note that, as mentioned before, the precise form of
the shape of $\tilde{z}(u)$, corresponding to the broadened delta
function \Eq{DeltaT}, {matters only in $0D$ regime}.

\

\begin{figure}[t]
  \ifpdf
    \includegraphics[width=\columnwidth]{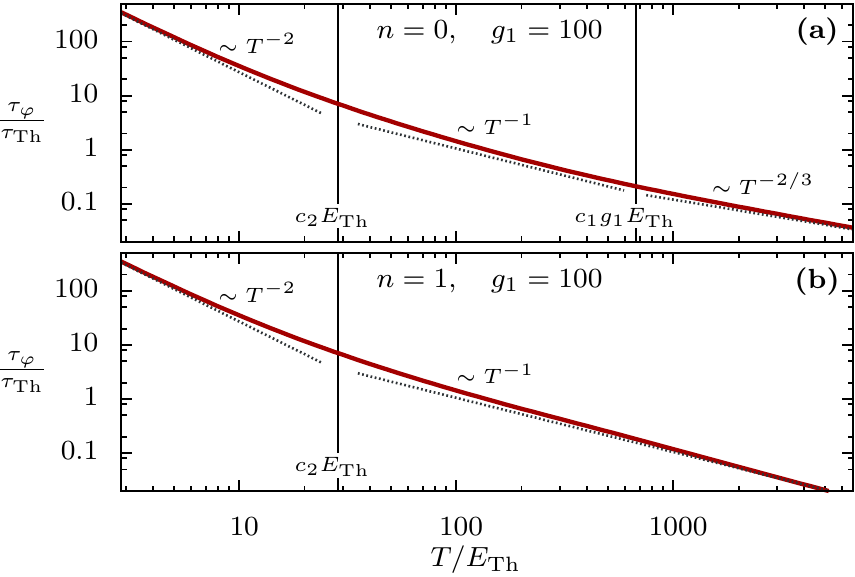}
  \else
    \epsfig{file=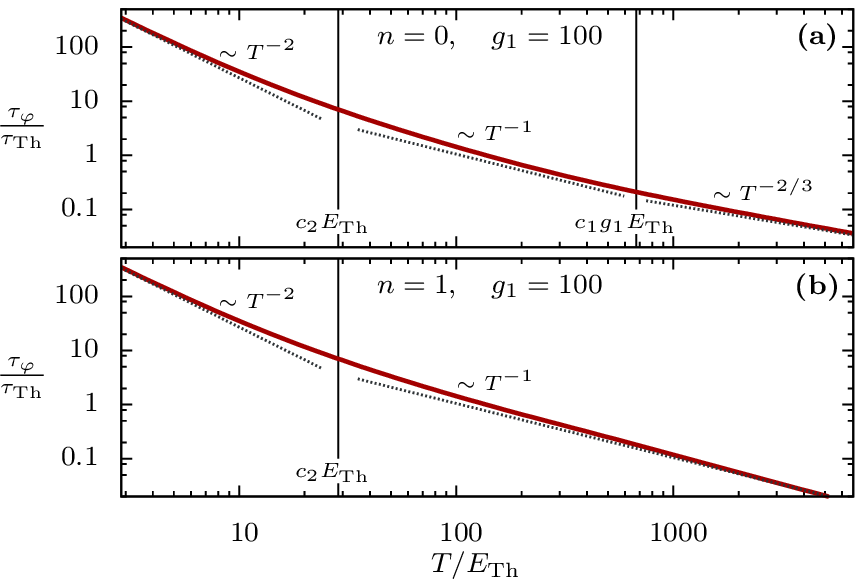,width=\columnwidth}
  \fi
  \label{FigDecayFunc}
  \caption{Dephasing time $\tPh$ extracted from \Eq{DecFuncFinal} and
           $\decayfunc_n(\tPh) = 1$ for $g_1 = 100$. (a) Shows the
           result for zero winding number $n=0$ and (b) for $n=1$. 
           For winding numbers $|n|>0$ the diffusive regime,
           $\tPh \sim T^{-2/3}$, is absent.}
\end{figure}

\noindent
To summarize, we found the following regimes:
\begin{equation}\nonumber\refstepcounter{equation}\label{LimCases}
  {\cal F}_n (t) \simeq
   \left\{
     \begin{array}{lclr}
       \displaystyle
       \frac{\pi^{3/2}}{2 g_1} \sqrt{\ETh}T t^{3/2} \quad , 
        & \!\!\! & \tT \ll t \ll \tTh, n=0; &
        \;\; (\theequation \text{a}) \\
       \displaystyle
       \frac{\pi T t}{3 g_1},
        & \!\!\! & \tT \ll t \ll \tTh, |n|>0; &
        \;\; (\theequation \text{b}) \\
       \displaystyle
       \frac{\pi T t}{3 g_1},
        & \!\!\! & \tT \ll \tTh \ll t, {\rm all \ } n; &
        \;\; (\theequation \text{c}) \\
       \displaystyle
       \frac{\pi^2}{270 \, g_1} \frac{T^2 t}{\ETh},
        & \!\!\! & \tTh \ll \tT \ll t, {\rm all \ } n. & 
        \;\; (\theequation \text{d})
     \end{array}
   \right.
\end{equation}
Note that the crossover temperatures where
$\tau_\varphi^{\rm diff} \simeq \tau_\varphi^{\rm erg}$ 
or $\tau_\varphi^{\rm erg} \simeq \tau_\varphi^{0D}$, 
namely  $c_1 g_1 \ETh$ or $c_2 \ETh$, respectively,
involve large  prefactors, $c_1 = 27/4 \simeq 7$
and $c_2 = 90/\pi \simeq 30$.
This can be seen in a numerical evaluation of \Eq{DecFuncFinal},
which is presented in Fig.~\ref{FigDecayFunc}.
In particular, one observes that the onset
of the $0D$ regime is already at temperatures smaller than
$30 \ETh$, i.e. well above $\ETh$. This should {significantly}
simplify experimental efforts to reach this regime.

\section{Correction to the conductance}

Inserting \Eq{CooperonSum} into \Eq{WLSigma}, we obtain the temperature
dependent correction to the conductance
\begin{align}\label{FinalConductance}
  \Delta g(T, \phi) = & -\frac{4 L}{g_1 \tTh} 
                        \int_0^{\infty} \intd{t}
                        \sum_{n=-\infty}^{+\infty} \\
                      & \qquad
                        \frac{e^{-(n/2)^2 \tTh/t}}{\sqrt{4 \pi D t}}
                        e^{ -t/\tH -t/\tDw -\decayfunc_n(t)}
                        \cos(4 \pi n \, \phi / \phi_0 )
  \, . \nonumber
\end{align}
The resulting value of $|\Delta g(T, \phi)|$ increases with decreasing
$T$ in a manner governed by $\tPh$. We recall that {in the high
temperature regime} dephasing can be relatively strong, {so that}
one can neglect effects of dissipation (i.e. particle escape out of
the ring) and of the external magnetic field on the Cooperon if
$\tPh(T) \ll {\rm min} [\tH, \tDw]$. In the diffusive regime,
$\tPh \ll \tTh$, $\Delta g(T, \phi)$ is dominated by the trajectories
with $n = 0$ since the contribution of the trajectories with
$|n| \geq 2\sqrt{t/\tTh} \sim \sqrt{\tPh/\tTh}$ is  exponentially small.
Thus we arrive at \cite{1982_aak_electroncollisions}:
\begin{equation}\label{ConductanceLimitDiffusive}
  |\Delta g| \simeq \frac{2}{g_1} \sqrt{\frac{\tPh}{\tTh}}
             \propto \left( \frac{\ETh}{g_1^2 T} \right)^{1/3}
  \, .
\end{equation}
In contrast, the trajectories with large winding number contribute
in the ergodic regime, $\tT \ll \tTh \ll \tPh$, therefore, converting
the sum to the integral
$\sum_n \exp(-(n/2)^2 \tTh/t)
 \simeq \int {\rm d}n \exp(-(n/2)^2 \tTh/t)
 \sim \sqrt{t/\tTh}$,
Eq.(\ref{FinalConductance}) yields  \cite{2004_ludwig_mirlin_ring,
2005_texier_montambaux_ring}
\begin{equation}\label{ConductanceLimitErgodic}
  |\Delta g| \simeq \frac{4}{g_1} \frac{\tPh}{\tTh}
             \propto \frac{\ETh}{T} \, .
\end{equation}

Dephasing due to electron interactions becomes \emph{weak} in the $0D$
regime and, therefore, the situation drastically changes at the 
crossover from the ergodic regime to the $0D$ one. In particular, we
find $\tPh \gg g_1 \tTh$, see Eq.(\ref{LimCases}c) and as far as $g_1$
is large, one may enter a low temperature regime where $\tPh \ge \tDw$.
In this  case, the temperature independent parts of the Cooperon
decay must be taken into account. In our model, with decreasing $T$,
the growth of $|\Delta g (T, \phi)|$ saturates towards 
$|\Delta g (0, \phi)|$ once $\tPh$ increases past $\min[\tH,\tDw]$
(a more quantitative  consideration is given in the next section).
Nevertheless, the temperature dependence of $\Delta g$ is still
governed by $\tPh(T)$ and we can single it out by subtracting the
conductance from its limiting value at $T=0$. Then the difference
\begin{equation}\label{ConductanceLimit0D}
  |\Delta g(0, \phi)| -|\Delta g(T, \phi)|
    \simeq \frac{4}{g_1} \frac{\tDw^2}{\tTh \tPh}
    \propto \left( \frac{\tDw T}{g_1} \right)^2
\end{equation}
shows $T^2$-behaviour in the  $0D$ regime.

\section{Suggested Experiments}

Our theoretical predictions {should be observable} in real
experiments, {provided that several requirements are met}. We list
these conditions in accordance with their physical causes, focusing
below on the example of a ring prepared from a quasi-$1D$ wire of width
$L_W$ on a $2D$ surface.

\subsection{Validity of theoretical predictions}\label{SectionValidity}

\paragraph{$1D$ diffusion:} 
We have used the theory of $1D$ diffusion which
calls for the following inequalities 
\begin{equation}
  {
  L \gg (\ell, L_W) \gg \lambda_{\text F} \, ;
  }
\end{equation}
$\lambda_{\text F}$ is the Fermi wavelength. 

\paragraph{Weak localization regime:}
Eq.(\ref{FinalConductance}) describes the leading weak localization
correction to the conductance. Subleading corrections can be neglected
if (a) the classical conductance of the ring is large
\begin{equation}
  g_1 \propto (\ell/L)(L_W/\lambda_{\text F}) \gg 1 \, ;
\end{equation}
and (b) the leading correction to the conductivity is smaller than its
classical value, $|\Delta g| < 1$. {The former condition can be
  assured by a} proper choice of the ring geometry and of the material
while, in the low temperature regime, the latter is provided by finite
dissipation.

\paragraph{Time-/spatial-dependence of the noise correlation function:}
We have used the noise correlation function \eqref{Vcorrelator} where
dependences on time- and space-coordinates are factorized. This
simplified form requires the following condition (see Section 1.3):
\begin{equation}\label{UnitaryLimitDenominator}
  2 \nu D \qcoord^2 \gg (D\qcoord^2 - i \omega)/V(\qcoord) \, .
\end{equation}
In the $0D$ regime we can roughly estimate typical values{,}
$D \qcoord^2 \sim \omega \sim \ETh${,} arriving at the inequality
\begin{equation}\label{UnitaryLimit0}
  \nu V(\qcoord) \gg 1 \, .
\end{equation}
For a quasi-$1D$ wire on a $2D$ structure, $\nu$ and $V$ can be written
as (restoring $\hbar$)
\begin{equation}\label{DOS2D}
  \nu_{2 {\rm D}} = \frac{m_e}{2 \pi \hbar^2} \, ,
\end{equation}
where $m_e$ is the electron mass, and
\begin{equation}\label{CoulombPot}
  V_{1 {\rm D}}(q) = \frac{e^2 L_W}{4 \pi \epsilon_0} |\ln(L_W^2 q^2)| \,
\end{equation}
(in SI units).
{Thus, \eqref{UnitaryLimit0} implies that $L_W$ cannot be taken to
be overly small. Inserting material parameters, however, this condition
turns out not to be very restrictive, as long as $\nu_{2 {\rm D}}$ is
reasonably large.}

\paragraph{Contacts (dissipation and absence of the Coulomb blockade):}
The {presence of contacts, through which electrons can escape into
leads, is mimicked in our model through the}
homogeneous dissipation rate
$1/\tDw$. We have assumed weak dissipation:
\begin{equation}
  \tTh \lesssim \tDw \, .
\end{equation}
This ensures that the winding trajectories with $|n| \ge 1$, responsible
for AAS oscillations, are relevant. On the other hand,
{$\tau_{\rm dw}$ cannot be taken to be arbitrarily large, since the
growth of the WL correction to the conductance with decreasing
temperature is cut off {mainly} due to this temperature independent
dissipation, and this cutoff has to occur sufficiently soon that the
relative correction remains small, else we would leave the WL regime.}
Choosing the zero temperature limit, somewhat arbitrary, as
$|\Delta g (0, \phi)| = 1/2$, we find from \Eq{ConductanceLimitErgodic}
\begin{equation}
  \tDw / \tTh \lesssim g_1/8 \, .
\end{equation}
We note that our assumptions imply $ \, \tPh \gg g_1 \tTh > \tDw \, $ in
the $0D$  regime, i.e., dephasing due to electron interactions is weak
(each electron contributing to transport through the ring is dephased
only a little bit during the course of its stay in the ring).
Nevertheless, we demonstrate below (see Fig.~\ref{FigConductance}) that 
the $ \, T^2 \, $-dependence of the conductance should be visible in
real experiments.

To choose a suitable value for the conductance at the contacts, we
estimate $\tDw / \tTh \simeq g_1 / \gcont$, which results in
\begin{equation}\label{GContLarge}
  8 \lesssim \gcont \, .
\end{equation}
We suppose that the contacts are open and have a maximal transmission
per  channel at the contact
\begin{equation}
  T_{\rm cont} = 1 \  \Rightarrow \  \gcont = T_{\rm cont} N = N \, ,
\end{equation}
($ \, N \, $ is the number of transmitting reflectionless channels at
the contact). This choice allows one to maximize the WL effect and,
simultaneously, to minimize any Coulomb blockade effects, which we have
neglected.

\subsection{Possible experimental setup}

\paragraph{Temperature range:}
The relevant temperature range, $[\Tmin, \Tmax]$, is limited from below
by dilution refrigeration $(\Tmin \simeq 10{\rm mK})$ and from above by
our neglect of phonons ($\Tmax \simeq 5$K). Furthermore, the ring should
be small enough that $c_2 \ETh \gtrsim \Tmin$; $ \, c_2 \ETh \, $ is the
upper estimate for the temperature of the crossover to the $0D$ regime,
see the discussion after \Eq{LimCases}.

\begin{figure}[t]
  \ifpdf
    \includegraphics[width=\columnwidth]{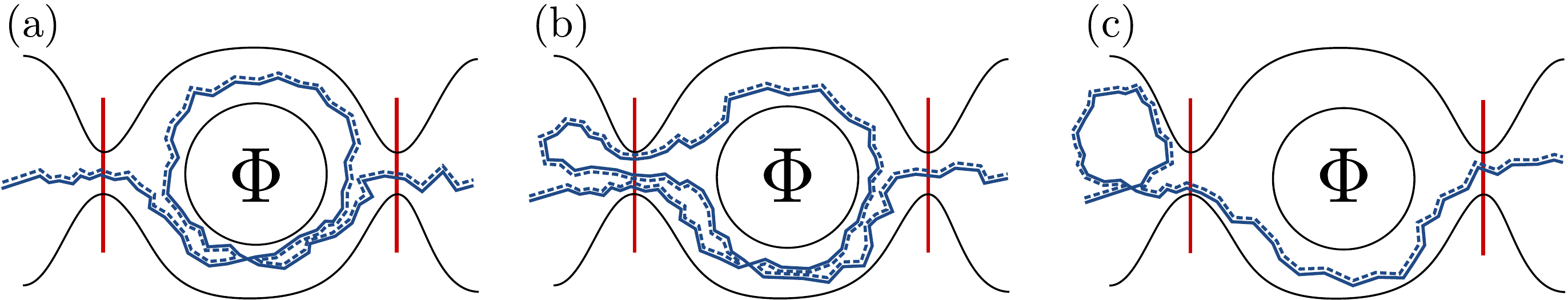}
  \else
    \epsfig{file=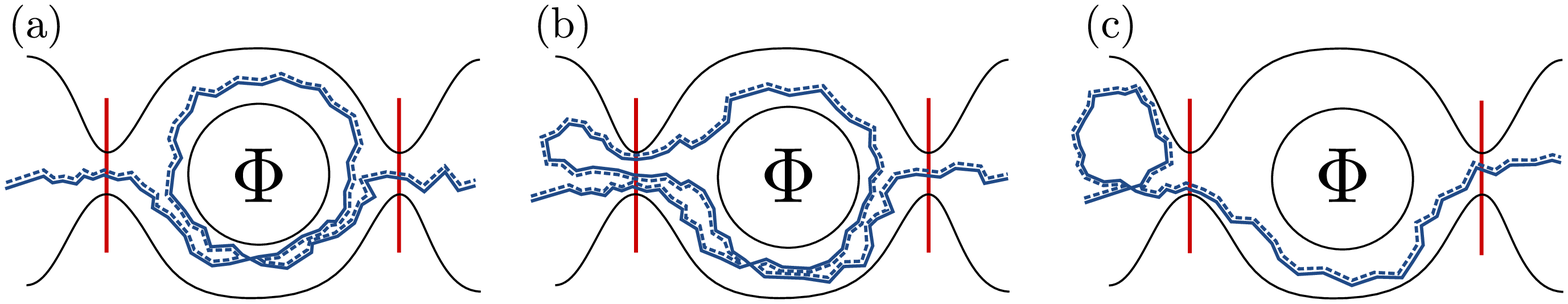,width=\columnwidth}
  \fi
  \caption{{(a) a "ring"-Cooperon, confined entirely to the ring; 
           (b) a "cross"-Cooperon, extending from the ring to the lead and back;
           (c) a "lead"-Cooperon, confined entirely to the leads.}}
\label{FigureRingContributions}
\end{figure}

\paragraph{Contributions from the leads:}
We have considered an ideal situation and calculated the Cooperon decay
function for the isolated ring, where the finite dissipation
rate $1/\tDw$ does not affect the decay function up to leading order in
$\tTh/\tDw$. This means that the Cooperons are assumed to live
completely inside the ring and not influenced by dephasing in the leads,
i.e. it corresponds to the situation shown in
Fig.~\ref{FigureRingContributions}(a).

In real experiments, the correction to the conductance, $\Delta g$, is
sensitive to dephasing in the leads because Cooperons exist which
either belong to the lead (e.g. the situation shown in
Fig.~\ref{FigureRingContributions}(c)) or extend over both the ring and
the lead (Fig.~\ref{FigureRingContributions}(b))
\cite{2006_blanter_vinokur_glazman_wlingranularmedia}{.}
(Note that in contrast to 
\Ref{2007_whitney_suppressionofweaklocalization} or
\Ref{2006_blanter_vinokur_glazman_wlingranularmedia},
we do not consider Cooperons with a Hikami-box directly at the contact,
since we chose $T_{\rm cont} = 1$.)

Contributions of such trajectories might mask the signatures of
dephasing in the confined region (the ring). This concern also applies
to quantum dots connected to leads (cf. the
$\tPh \propto T^{-1}$-behavior observed in
\Refs{1998_huibers_opendot1,1998_huibers_opendot2,
1999_huibers_opendot3}), or finite-size effects in a network of
disordered wires \cite{2008_ferrier_grid}, where paths encircling a
given unit cell might spend significant time in neighboring unit
cells as well (cf. $ T^{-1/3} $-behavior observed in
\Ref{2008_ferrier_grid} at $ \tPh / \tTh \geq 1 $).

We will now argue that if the lead dimensionless conductance is larger
than the contact conductance \cite{2009_brouwer_privatecomm}
\begin{equation}
  \glead \gg \gcont = N \, ,
\end{equation}
then the ring-Cooperon yields the dominating contribution to the WL
corrections.
{
Let us focus on the ergodic and 0D regimes, for which $\tPh \gg \tTh$,
so that that the Cooperon ergodically explores the entire ring. Then
the probability to find a closed loop in the ring is proportional to the
dwell time, $p_{\rm ring} \propto \tDw/\nu$, which is
$\propto 1/\gcont$. Thus we can estimate:
}
\begin{itemize}
  {
  \item the probability to enter the ring as
        $ \, p_{\rm in} \sim \gcont / \glead$; 
  \item the probability to find a closed loop in the ring as \\
        $ \, p_{\rm ring} \sim (\tDw/\nu) \sim 1 / \gcont $;
  \item the probability to exit the ring as
        $ \, p_{\rm out} \sim (\tDw/\nu) \gcont \sim 1 $; 
  \item the probability to find a closed loop in the diffusive lead as \\
        $ \, p_{\rm lead} \sim 1 / \glead $
  }
\end{itemize}
Using these estimates, the probabilities to find a ring-, cross-, or
lead-Cooperon are
\begin{eqnarray}
  P_{\rm C-ring}  & \sim & p_{\rm in}      \times p_{\rm ring}
                                           \times p_{\rm out}
                    \sim 1/\glead \, ; \\
  P_{\rm C-cross} & \sim & p_{\rm in}      \times p_{\rm ring}
                                           \times p_{\rm out}
                                           \times p_{\rm in}
                                           \times p_{\rm out}
                    \sim \gcont/\glead^2 \, ; \\
  P_{\rm C-lead}  & \sim & p_{\rm lead}    \times p_{\rm in}
                                           \times p_{\rm out}
                    \sim \gcont/\glead^2 \, ,
\end{eqnarray}
respectively. Thus we arrive at:
\begin{equation}\label{ProbEst}
  {
  P_{\rm C-lead} \sim P_{\rm C-cont}
  \sim P_{\rm C-ring} \times \gcont / \glead \ll P_{\rm C-ring} \, ,
  }
\end{equation}
which proves that the ring-Cooperon dominates the WL correction
for our choice of parameters if {$ \, \glead \gg \gcont$}. 

Since the $0D$ regime implies weak dephasing it is highly desirable to
improve the ``signal-to-noise'' ratio by filtering out contributions
which do not show $0D$ dephasing. This can be done
\cite{2008_ferrier_grid} by constructing from
$|\Delta g (T, \phi)|$ its nonoscillatory envelope
$|\Delta g_\envelope (T, \phi)|$, obtained by setting $\theta = 0$ in
\Eq{CooperonSum} while retaining $\tH \neq0$, and studying the
difference
\begin{equation}\label{Subtr}
  \Delta \overline{g}(T,\phi) 
    = | \Delta g_\envelope (T,\phi)| - | \Delta g (T,\phi)| \, .
\end{equation}
This procedure is illustrated in Fig.~\ref{FigureAASOsc}. The
lead-Cooperons do not have the Aharonov-Bohm phase and are eliminated by
this filtering procedure. Unfortunately, cross-Cooperons cannot be
filtered {in this manner, since they do experience the Aharonov-Bohm
phase.}
Nevertheless, if the condition {$\gcont \ll \glead$} holds,
$\Delta \overline{g}$ is completely dominated by paths residing only in
the ring in accordance with the estimate \Eq{ProbEst}.

\subsection{Numerical results for $2D$ GaAs/AlGaAs heterostructures}
  \label{SectionNumericalResults}

\begin{figure}[t]
  \ifpdf
    \includegraphics[width=\columnwidth]{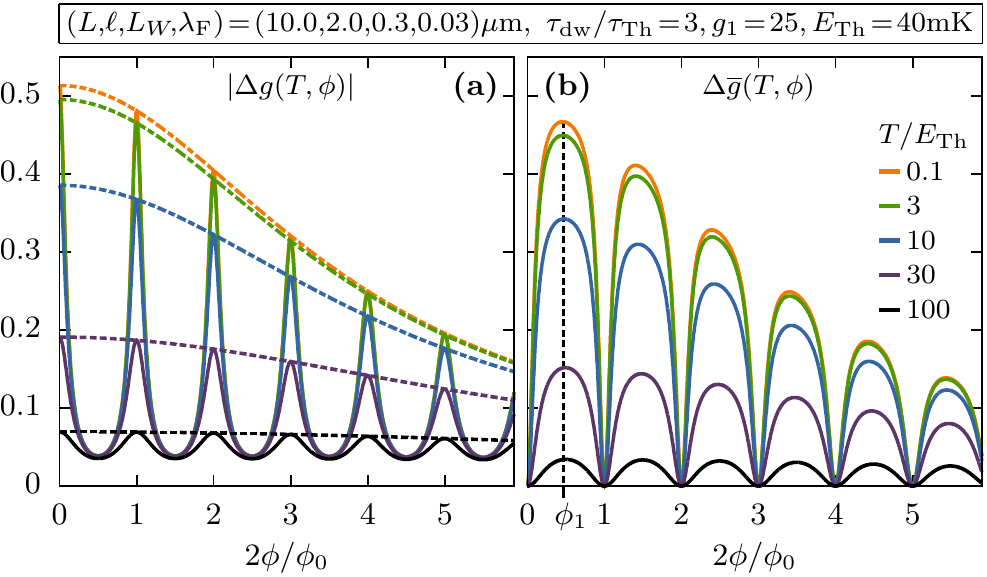}
  \else
    \epsfig{file=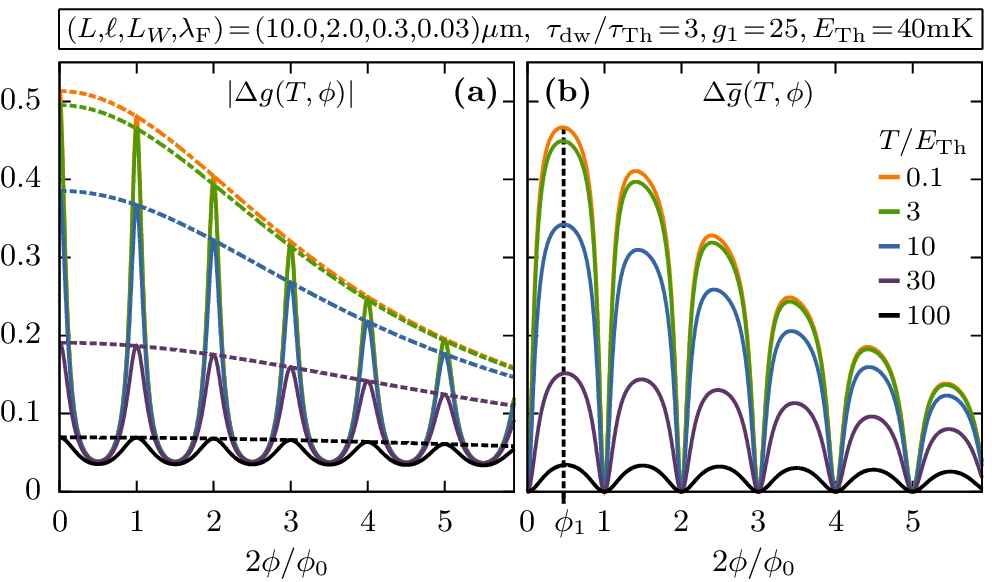,width=\columnwidth}
  \fi
  \caption{(a) The WL correction $ |\Delta g (T, \phi)|$
    (solid lines), its envelope $|\Delta g_\envelope (T, \phi)|$ (dashed
    lines) and (b) their difference
    $\Delta \overline{g} = |\Delta g_\envelope| - |\Delta g| $, plotted
    as function of magnetic flux $ 2\phi/\phi_0 $, for five different
    temperatures between $0.1 \ETh$ and $100 \ETh$ (increasing from top
    to bottom).}
\label{FigureAASOsc}
\end{figure}

\noindent
All above-mentioned constraints can be met, e.g., with rings prepared
from a $2D$ GaAs/AlGaAs heterostructure. In such systems, diffusive
behavior emerges from specular boundary scattering of the electrons,
see \Ref{1988_beenakker_vanhouten_boundary}, leading to the
following dephasing time due to the external magnetic field:
\begin{equation}
  \tH = 9.5 (c / e H )^2 \times (l/D L_W^3) \, .
\end{equation}
Furthermore, inserting \Eq{DOS2D} into the $2D$ conductivity
$\sigma_{2 {\rm D}} = 2 e^2 \nu_{2 {\rm D}} D$ with $D=v_{\rm F} \ell$,
we obtain the corresponding dimensionless conductance:
\begin{equation}
  g_1 = \frac{h}{e^2} \frac{\sigma_{2 {\rm D}} L_W}{L}
      = 4 \pi \frac{L_W \ell}{\lambda_F L} \, .
\end{equation}
A typical Fermi wavelength in a GaAs/AlGaAs heterostructure is
$\lambda_{\text F} \approx 30$nm ($v_F  \approx 2.5 \cdot 10^{5}$m/s)
\cite{2009_niimi_mesoscopicwire,1993_marcus_ballisticdot,
2008_ferrier_grid,2000_yevtushenko_antidots}.
Thus, by suitably choosing $L$, $L_W$ and $\ell$ we can adjust $g_1$
and $\ETh$ to make all regimes of the dephasing time accessible.

\begin{figure}[t]
  \ifpdf
    \includegraphics[width=\columnwidth]{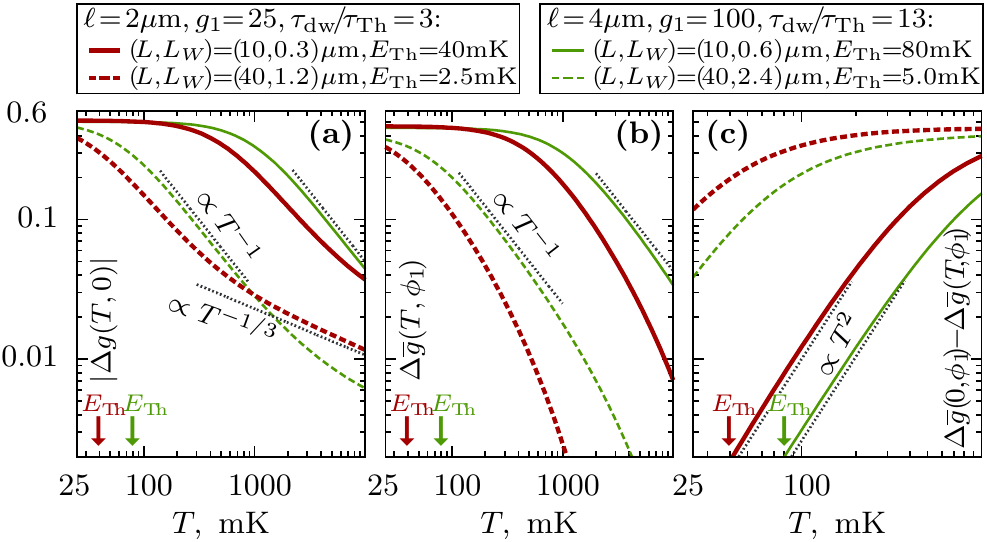}
  \else
    \epsfig{file=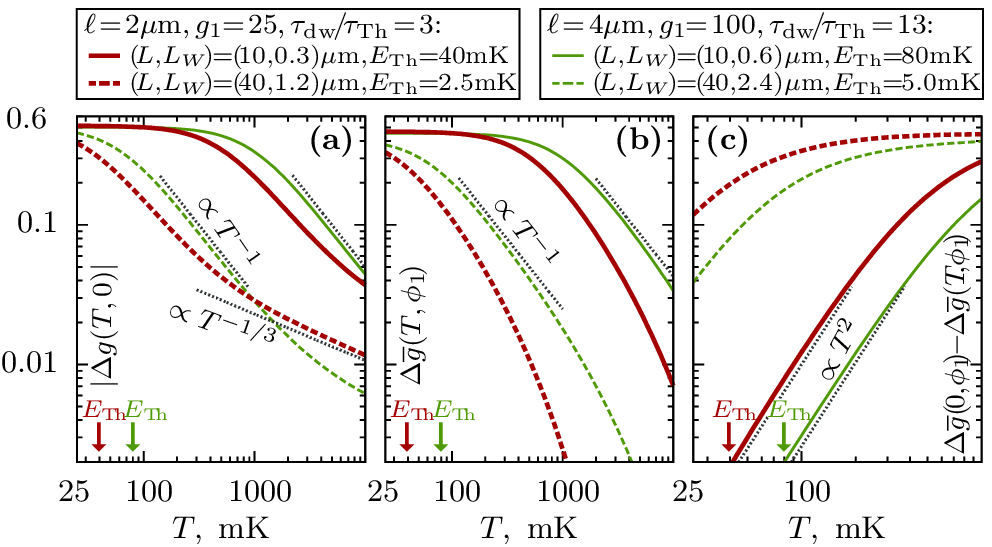,width=\columnwidth}
  \fi
\caption{$T$-dependence of (a) the WL correction at zero
         field, $ |\Delta g (T, 0)|$ and (b) at finite field
         with envelope subtracted, $\Delta\overline{g} (T, \phi_1)$;
         (c) the difference
         $\Delta\overline{g}(0,\phi_1) - \Delta\overline{g}(T,\phi_1)$,
         which reveals a crossover to $T^2$-behavior for
         $T \ll 30 \ETh$. The flux $ \phi_1 $, which weakly depends on
         $T$, marks the first maximum of $\Delta\overline{g}(T,\phi)$,
         see inset of Fig.~\ref{FigureAASOsc}.
         [This figure is reproduced from \Ref{2009_our_prb}]}
\label{FigConductance}\vspace{-2mm}
\end{figure}

Numerical results for $|\Delta g|$ and $\Delta \overline g$, obtained
from Eq.(\ref{WLSigma}) using experimentally realizable parameters,
are shown in Figs.~\ref{FigureAASOsc} and~\ref{FigConductance} for
several combinations of these parameters.
The regime where $\Delta g$ exhibits diffusive $T^{-1/3}$ behavior 
($7 g_1 \ETh \ll T \ll \Tmax$) is visible only for our smallest
choices of both $g_1$ and $\ETh$ (Fig.~\ref{FigConductance}(a), heavy
dashed line). AAS oscillations in $|\Delta g|$ and $\Delta \overline g$
(Fig.~\ref{FigureAASOsc}), which require $\tTh \ll \tPh$, first emerge
at the crossover from the diffusive to the ergodic regime. They increase
in magnitude with decreasing $T$, showing ergodic $T^{-1}$ behavior for
$30 \ETh \ll T \ll 7 g_1 \ETh$ (Figs.~\ref{FigConductance}(a),(b)), and
eventually saturate towards their $T=0$ values, with
$\Delta\overline{g}(0,\phi) - \Delta\overline{g}(T,\phi)$
showing the predicted $0D$ behavior, $\propto T^2$, for
$T \lesssim 5 \ETh$, see Fig.~\ref{FigConductance}(c).

\section{Conclusions}

For an almost isolated disordered quasi-$1D$ ring with $T \gg \ETh $,
the $T$-dependence of the dephasing time has been known to behave as
$\tPh \propto T^{-2/3}$ \ (\Ref{1982_aak_electroncollisions}) or
$\propto T^{-1}$ (\Refs{2004_ludwig_mirlin_ring,
2005_texier_montambaux_ring}) in the diffusive or ergodic regimes,
respectively.
Here we showed how it crosses over, for $T \ll  30 \ETh$, to
$\tPh \propto T^{-2}$, in agreement with the theory of dephasing in $0D$
systems (\Ref{1994_sivan_imry_aronov_quantumdot}).
This crossover manifests itself in both the smooth part of the
magnetoconductivity and the amplitude of the AAS oscillations.
Importantly, the latter fact can be exploited to decrease the effects
of dephasing in the leads, by subtracting from the magnetoconductivity
its smooth envelope.
While we did not give an exhaustive study of all contributions to
dephasing in the connected ring, we were able to show
that the leading contribution results only from trajectories confined to
the ring.
Thus, an analysis of the $T$-dependence of the AAS oscillation amplitude
may offer a way to finally observe, for $T \lesssim 5 \ETh$, the elusive
but fundamental $0D$ behavior $\tPh \sim T^{-2}$. Its observation,
moreover, would allow \emph{quantitative} experimental tests of the
role of temperature as ultraviolet frequency cutoff in the theory of
dephasing.
An interesting challenge for future works consists in {a} more
realistic model of the connection to the leads. Work on the model of an
$N$-channel ring attached via two arms with fewer channels to absorbing
boundaries is currently in progress \cite{2010_tobepublished}.

\acknowledgments We acknowledge illuminating discussions with
B.~L.~Altshuler, C.~B\"auerle, N.~O.~Birge, Ya.~M.~Blanter,
P.~W.~Brouwer, L.~I.~Glazman, Y.~Imry, V.~E.~Kravtsov, J.~Kupferschmidt,
A.~D.~Mirlin, Y.~V.~Nazarov, A.~Rosch, D.~Weiss and V.~I.~Yudson.
{We acknowledge}
support from the DFG through SFB TR-12, the Emmy-Noether program
and the Nanosystems Initiative Munich Cluster of Excellence;
from the NSF, Grant No. PHY05-51164;
and from the EPSRC, Grant No. T23725/01.

\bibliographystyle{ws-rv-van}

\printindex                         % to print subject index
\end{document}